\documentclass[12pt]{article}
\usepackage{epsf}
\usepackage[dvips]{graphicx,psfrag}

\renewcommand{\thefootnote}{\#\arabic{footnote}}
\begin{document}

\newcommand{\gtrsim}{ \mathop{}_{\textstyle \sim}^{\textstyle >} }
\newcommand{\lesssim}{ \mathop{}_{\textstyle \sim}^{\textstyle <} }

\renewcommand{\thefootnote}{\fnsymbol{footnote}}
\setcounter{footnote}{0}
\begin{titlepage}

\def\thefootnote{\fnsymbol{footnote}}

\begin{center}

\hfill TU-671\\
\hfill IFP-799-UNC\\
\hfill hep-ph/0210392\\
\hfill October, 2002\\

\vskip .5in

{\Large \bf

Cosmic Microwave Background \\ from Late-Decaying
Scalar Condensations\footnote
{Talk given by T. Moroi at SUSY02, ``The 10th International Conference
on Supersymmetry and Unification of Fundamental Interactions,'' DESY,
Hamburg, June 17-23, 2002.}

}

\vskip .45in

{\large
Takeo Moroi$^{(a)}$ and Tomo Takahashi$^{(a,b)}$
}

\vskip .45in

$^{(a)}$
{\em
Department of Physics, Tohoku University, Sendai 980-8578, Japan
}

\vskip .2in

$^{(b)}$
{\em
Department of Physics and Astronomy,
University of North Carolina\\
Chapel Hill, NC 27599-3255
}

\end{center}

\vskip .4in

\begin{abstract}

    We study the cosmic microwave background (CMB) anisotropy in the
    scenario with late-decaying scalar condensations which arise in
    many class of cosmological scenarios based on supersymmetric
    models.  With such a scalar condensation $\phi$, the CMB radiation
    we observe today originates to $\phi$ and hence the CMB anisotropy
    is affected if $\phi$ has a primordial fluctuation.  In
    particular, if all the components in the universe (i.e.,
    radiation, baryon, cold dark matter, and so on) are generated from
    the decay product of $\phi$, the dominant source of the cosmic
    density perturbations can be the primordial fluctuation of $\phi$,
    not the fluctuation of the inflaton field.  In this case, the
    constraints on the inflation models can be drastically relaxed.
    In other case, the baryon or the CDM may not be from the decay
    product of $\phi$ and correlated mixture of the adiabatic and
    isocurvature fluctuations can be generated.  If so, the CMB
    angular power spectrum may not be the same as the adiabatic result
    and the on-going MAP experiment may observe a deviation from the
    prediction of the standard inflationary scenario.

\end{abstract}

\end{titlepage}

\renewcommand{\thepage}{\arabic{page}}
\setcounter{page}{1}
\renewcommand{\thefootnote}{\#\arabic{footnote}}
\setcounter{footnote}{0}

In supersymmetric models, it is known that there are various light
scalar fields which play important roles in cosmology.  In particular,
in many class of scenarios, there exists a scalar field (other than
the inflaton) which once dominates the universe and decays at a later
stage of the evolution of the universe.  Indeed, this is the case in,
for example, Affleck-Dine baryogenesis \cite{NPB249-361} and
sneutrino-induced leptogenesis \cite{snu-leptogen}.  In addition, if
the modulus field acquires a large amplitude at the beginning of the
universe, it is expected to decay after the big-bang nucleosynthesis
(BBN) and spoils the success of the BBN if their masses are of $O(100\ 
{\rm GeV} - 1\ {\rm TeV})$ \cite{PRD63-103502}.  One solution to this
cosmological difficulty is to push up the mass of the modulus field
and make it decay before the BBN starts \cite{heavy-moduli}.  Although
these scenarios have been attracted many attentions, it is very
difficult to observe the consequence of the late-decaying scalar
condensations.

One of the most important consequences of such scenarios is that the
cosmic microwave background (CMB) radiation we observe today
originates to the late-decaying scalar field (denoted as $\phi$
hereafter) rather than the inflaton field.  Thus, if there exists a
primordial fluctuation in the amplitude of $\phi$, it becomes a new
source of the cosmic density perturbations and affects the CMB
anisotropy.  In particular, recently the observation of the CMB power
spectrum is being greatly improved and hence it may be possible to
observe some signal in the CMB anisotropy from the late-decaying
scalar condensations.  Thus, it is very important to discuss the CMB
anisotropy taking account of the effects of the late-decaying scalar
condensations, which is the subject of the study here.  We will see
that, if the scalar condensation acquires a primordial fluctuation, it
may significantly affect the CMB angular power spectrum.  In
particular, we will emphasize that the conventional constraints on the
inflation models can be greatly relaxed if such a scalar field exists.
In addition, in some case, correlated mixture of the adiabatic and
isocurvature density fluctuations can be generated in such a scenario
and the on-going MAP \cite{MAP} experiment may observe the signal of
the late-decaying scalar condensations.

Let us start our discussion with introducing the scenario we have in
mind.  Here, we consider the thermal history with a scalar field
$\phi$ which decays and produces a large amount of entropy at a late
stage of the evolution of the universe.  In this study, we assume
inflation to solve horizon, flatness, and other cosmological problems.
Then, after the inflation, the universe is reheated and the radiation
dominated universe is realized.  We call this epoch as the RD1 epoch.

If the potential of $\phi$ is dominated by the parabolic term, $\phi$
changes its behavior depending on the relative size of $H$ and
$m_\phi$, where $m_\phi$ is the mass of $\phi$ and $H$ is the
expansion rate of the universe.  In the very early universe, the total
energy density of the universe is extremely large and hence the
relation $H\gg m_\phi$ holds.  In this case, the slow-roll condition
is satisfied and $\phi$ takes (almost) constant value.  On the
contrary, as the universe expands, $H$ becomes smaller than $m_\phi$
at some point.  Then, $\phi$ oscillates and the energy density of
$\phi$ decrease as $\rho_\phi\propto a^{-3}$.  Consequently, the
energy density of the radiation decreases faster than $\rho_\phi$ if
$H\lesssim m_\phi$.  Thus, if the initial amplitude of the scalar
field, denoted as $\phi_{\rm init}$, is large enough, the universe is
once dominated by the scalar field $\phi$ after the RD1 epoch.  (We
call this epoch as $\phi$D epoch.)  In the following, we assume that
$\phi_{\rm init}$ is large enough so that the $\phi$D epoch is
realized.  Then, $\phi$ decays when $H\sim\Gamma_\phi$ with
$\Gamma_\phi$ being the decay rate of $\phi$.  After the decay of
$\phi$, the radiation-dominated universe is realized.  We call this
epoch as RD2 epoch.

In this framework, the CMB radiation we observe today originates to
the scalar field $\phi$, not to the inflaton field.  Thus, if $\phi$
has an primordial fluctuation, it also affects the CMB anisotropy.
Such a fluctuation is expected to be generated during the inflation;
for the Fourier mode with comoving momentum $k$, the initial value of
the fluctuation of $\phi$ is given by $\delta\phi (t, \vec{k}) =
\left[ {H_{\rm inf}}/{2\pi} \right]_{k=aH_{\rm inf}}$, where
$H_{\rm inf}$ is the expansion rate of the universe during the
inflation.  (Here, we assumed $m_\phi\ll H_{\rm inf}$.)

In this class of scenario, there are two independent sources of the
cosmic density perturbations; one is the primordial fluctuation of the
inflaton field $\chi$ and the other is the primordial fluctuation of
the scalar field.  Since we use the linear perturbation theory, their
effects can be discussed separately.  The effect of the inflaton-field
fluctuation $\delta\chi$ is parameterized by the metric perturbation
$\Psi^{(\delta\chi)}$ generated by the inflaton fluctuation.  (The
superscript $(\delta\chi)$ is for perturbations from the inflaton
fluctuation.)  It is well-known that inflaton fluctuation provides the
adiabatic density fluctuations, i.e., no entropy perturbation is
generated from $\delta\chi$.  Thus, in the following, we study the
effects of the primordial fluctuation of the scalar field $\phi$.

Since we are interested in the CMB anisotropy, it is crucial to
understand the relations among the density fluctuations of each
components, like photon, CDM, baryon, and so on, after the decay of
the scalar field $\phi$.  For this purpose, we define the perturbed
line element (in the Newtonian gauge) as
\begin{eqnarray}
    ds^2 &=& - (1 + 2\Psi) dt^2
    + a^2 (1 + 2\Phi) \delta_{ij} dx^i dx^j.
\end{eqnarray}
In addition, from the density fluctuations of each components in the
Newtonian gauge, we define
\begin{eqnarray}
    \delta_X = \delta\rho_X / \rho_X,
\end{eqnarray}
where $X=r$, $c$, and $b$, corresponding to radiation,\footnote
{We assume that there is no entropy between the photon and the
neutrinos.}
CDM, and baryon, respectively.

To parameterize the density fluctuations of each components after the
decay of $\phi$, it is convenient to consider the ``entropy'' between
the $\phi$ field and the components generated from the decay product
of the inflaton field before the decay of $\phi$:
\begin{eqnarray}
    S_{\phi\chi}^{(\delta\phi)} (k) = 
    \frac{\delta\rho_\phi (t, k)}{\rho_\phi (t)} - \delta_\chi (t)
    = \frac{2\delta\phi_{\rm init} (k)}{\phi_{\rm init}},
    \label{s_pc}
\end{eqnarray}
where $\delta\phi_{\rm init}$ is the initial fluctuation of $\phi$ and
the superscript ``$(\delta\phi)$'' is for perturbations generated from
the primordial fluctuation of $\phi$.

Density (and other) fluctuations in the RD2 epoch are generally
parameterized by using $S_{\phi\chi}^{(\delta\phi)}$.  If a component
$X$ is generated from the decay product of $\phi$, then there is no
entropy between the photon and $X$.  On the contrary, if the component
$X$ has some source other than $\phi$, the entropy between the photon
and $X$ is the same as $S_{\phi\chi}^{(\delta\phi)}$.  Thus, if all
the components in the universe are generated from $\phi$, the density
fluctuations become purely adiabatic and
\begin{eqnarray}
    \left[ \delta_\gamma^{(\delta\phi)} \right]_{\rm RD2} 
    = \frac{4}{3}\left[ \delta_b^{(\delta\phi)} \right]_{\rm RD2}
    = \frac{4}{3}\left[ \delta_c^{(\delta\phi)} \right]_{\rm RD2}
    = -2 \Psi_{\rm RD2}^{(\delta\phi)},
\end{eqnarray}
where the subscripts $\gamma$, $b$, and $c$ are for the photon,
baryon, and CDM, respectively.  In this case, the isocurvature
perturbation in the $\phi$ field is converted to the purely adiabatic
density perturbation after the decay of $\phi$
\cite{MoroiTakahashi,NPB626-395,PLB524-5}.  On the contrary, if the
baryon asymmetry does not originate to the decay product of $\phi$,
the entropy between the radiation and the baryon becomes
$S_{\phi\chi}^{(\delta\phi)}$ and hence \cite{MoroiTakahashi}
\begin{eqnarray}
    \left[ \delta_\gamma^{(\delta\phi)} \right]_{\rm RD2} 
    = \frac{4}{3}\left[ \delta_c^{(\delta\phi)} \right]_{\rm RD2}
    = -2 \Psi_{\rm RD2}^{(\delta\phi)},~~~
    \left[ \delta_b^{(\delta\phi)} \right]_{\rm RD2}
    = \frac{3}{4} \left[ \delta_\gamma^{(\delta\phi)} \right]_{\rm RD2} 
    + \frac{9}{2} \Psi_{\rm RD2}^{(\delta\phi)},
    \label{Sb}
\end{eqnarray}
and in the case where all the components other the CDM are generated
from the decay product of $\phi$,
\begin{eqnarray}
    \left[ \delta_\gamma^{(\delta\phi)} \right]_{\rm RD2} 
    = \frac{4}{3}\left[ \delta_b^{(\delta\phi)} \right]_{\rm RD2}
    = -2 \Psi_{\rm RD2}^{(\delta\phi)},~~~
    \left[ \delta_c^{(\delta\phi)} \right]_{\rm RD2}
    = \frac{3}{4} \left[ \delta_\gamma^{(\delta\phi)} \right]_{\rm RD2} 
    + \frac{9}{2} \Psi_{\rm RD2}^{(\delta\phi)}.
    \label{Sc}
\end{eqnarray}
In addition, if the baryon and the CDM are both generated from sources
other than $\phi$, we obtain
\begin{eqnarray}
    \left[ \delta_\gamma^{(\delta\phi)} \right]_{\rm RD2} 
    = -2 \Psi_{\rm RD2}^{(\delta\phi)},~~~
    \left[ \delta_b^{(\delta\phi)} \right]_{\rm RD2}
    = \left[ \delta_c^{(\delta\phi)} \right]_{\rm RD2}
    = \frac{3}{4} \left[ \delta_\gamma^{(\delta\phi)} \right]_{\rm RD2} 
    + \frac{9}{2} \Psi_{\rm RD2}^{(\delta\phi)}.
    \label{Sbc}
\end{eqnarray}
It is important to notice that, for the cases given in Eqs.\ 
(\ref{Sb}) $-$ (\ref{Sbc}), the entropy perturbations are correlated
with the adiabatic perturbation.\footnote
{Here, we assumed that the baryon number asymmetry and/or the CDM is
generated before the $\phi$ field starts to oscillate.  In other case,
the size of the correlated entropy perturbation may change.  For
details, see \cite{MoroiTakahashi}.}

Now, we are at the point to discuss the CMB anisotropy.  The CMB
anisotropy is characterized by the angular power spectrum $C_l$ which
is defined as
\begin{eqnarray}
    \left\langle \Delta T(\vec{x}, \vec{\gamma}) 
        \Delta T(\vec{x}, \vec{\gamma}')  \right\rangle_{\vec{x}}
    = \frac{1}{4\pi} 
    \sum_l (2l+1) C_l P_l (\vec{\gamma} \cdot \vec{\gamma}'),
\end{eqnarray}
with $\Delta T (\vec{x}, \vec{\gamma})$ being the temperature
fluctuation of the CMB radiation pointing to the direction
$\vec{\gamma}$ at the position $\vec{x}$ and $P_l$ is the Legendre
polynomial.  As we mentioned, there are two sources of the density
perturbations; the primordial fluctuations of the scalar fields $\chi$
and $\phi$.  Since there is no correlation between these fields, the
CMB anisotropies from these fluctuations are uncorrelated and the
resultant CMB power spectrum is given in the form
\begin{eqnarray}
    C_l 
    = C_l^{(\delta\chi)} + C_l^{(\delta\phi)},
    \label{Cl(tot)}
\end{eqnarray}
where $C_l^{(\delta \chi)}$ and $C_l^{(\delta \phi)}$ are
contributions from the primordial fluctuations of the inflaton field
$\chi$ and the $\phi$ field, respectively.  The inflaton contribution
$C_l^{(\delta\chi)}$ is known to be the adiabatic result.

The new contribution $C_l^{(\delta\phi)}$ depends on the properties of
the density perturbations of each components.  If all the components
in the universe (i.e., the photon, baryon, CDM, neutrino, and so on)
are dominantly generated from the decay product of $\phi$, there is no
entropy between any of two components.  In this case,
$C_l^{(\delta\phi)}$ is from adiabatic perturbations.  This fact has
an important implication.  In general, the scale dependences of
$\Psi^{(\delta\chi)}$ and $S_{\phi\chi}^{(\delta\phi)}$ are different.
In the slow-roll inflation scenario, $\Psi^{(\delta\chi)}$ is given by
\cite{PRD28-629}
\begin{eqnarray}
\Psi^{(\delta\chi)}_{\rm RD2} = \frac{4}{9}
\left[ \frac{H_{\rm inf}}{2\pi}\frac{3H_{\rm inf}^2}{V_{\rm inf}'} 
\right]_{k=aH_{\rm inf}},
\label{Psi(dchi)_RD2}
\end{eqnarray}
where $V_{\rm inf}'\equiv\partial V_{\rm inf}/\partial\chi$ with
$V_{\rm inf}$ being the inflaton potential.  On the contrary,
$S_{\phi\chi}^{(\delta\phi)}$ is related to the fluctuation of the
amplitude of $\phi$, as seen in Eq.\ (\ref{s_pc}), and hence
\begin{eqnarray}
S_{\phi\chi}^{(\delta\phi)} = 
\frac{2}{\phi_{\rm init}} 
\left[ \frac{H_{\rm inf}}{2\pi} \right]_{k=aH_{\rm inf}}.
\label{S(dphi)}
\end{eqnarray}
In many models of slow-roll inflation, the expansion rate $H_{\rm
inf}$ is almost constant during the inflation.  On the contrary, the
slope of the inflation potential $V_{\rm inf}'$ may significantly
vary.  As a result, $S_{\phi\chi}^{(\delta\phi)}$ becomes (almost)
scale independent while $\Psi^{(\delta\chi)}$ may have significant
scale dependence.  Since the currently measured CMB power spectrum
suggests (almost) scale invariant primordial density perturbation in
the conventional scenario, inflation models are excluded if
$\Psi^{(\delta\chi)}$ has too strong scale dependence
\cite{wang_et_al_2001}.

If the $\phi$ field exists, however, the situation may change
\cite{MoroiTakahashi}.  Since $S_{\phi\chi}^{(\delta\phi)}$ is
expected to be (almost) scale invariant, we can relax the constraint
on the inflation models if $C_l^{(\delta\phi)}$ becomes significantly
large, which happens when
$S_{\phi\chi}^{(\delta\phi)}\gtrsim\Psi^{(\delta\chi)}$.  As shown in
Eq.\ (\ref{s_pc}), $S_{\phi\chi}^{(\delta\phi)}$ is inversely
proportional to $\phi_{\rm init}$.  Thus, if the initial amplitude of
$\phi$ is small, this may happen and the CMB power spectrum may become
consistent with the observational data in a larger class of models of
inflation.

\begin{figure}[t]
    \begin{center}
        \scalebox{1.0}{\includegraphics{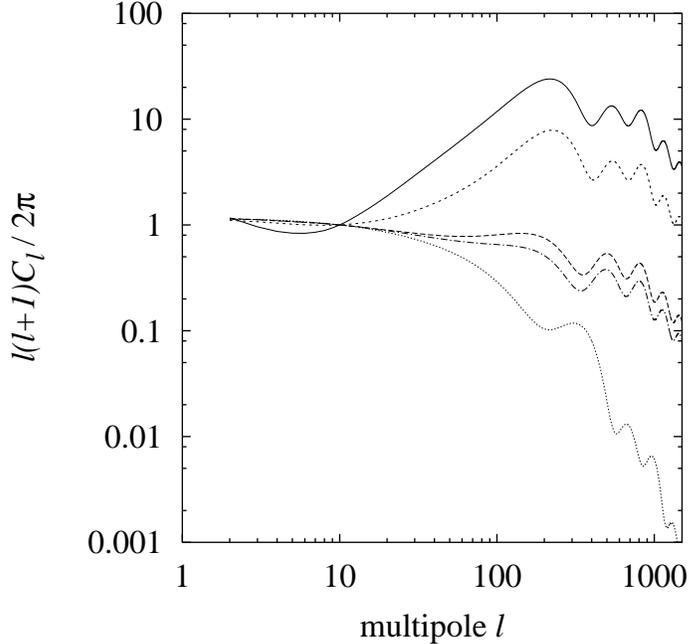}}
        \caption{The angular power spectrum with correlated mixture 
        of the adiabatic and isocurvature perturbations in the
        baryonic sector (solid line), in the CDM sector (long-dashed
        line), and in the baryonic and CDM sectors (dot-dashed line).
        (See Eqs.\ (\ref{Sb}), (\ref{Sc}), and (\ref{Sbc}),
        respectively.)  We also show the CMB angular power spectrum in
        the purely adiabatic (short-dashed line) and isocurvature
        density perturbations (dotted line).  We consider the flat
        universe with $\Omega_bh^2=0.019$, $\Omega_m=0.3$, and
        $h=0.65$, where $\Omega_b$ and $\Omega_c$ are the (present)
        density parameters for the baryon and the CDM, respectively,
        and $h$ is the Hubble constant in units of 100 km/sec/Mpc.
        The overall normalizations are taken as
        $[l(l+1)C_l/2\pi]_{l=10}=1$.}
        \label{fig:Cl's}
    \end{center}
\end{figure}

Now, we study effects of the correlated entropy fluctuations.  We
first plot the angular power spectrum with the correlated mixture of
the adiabatic and isocurvature perturbations in the baryonic and/or
CDM sector, i.e., the cases with the relations given in Eqs.\ 
(\ref{Sb}) $-$ (\ref{Sbc}).  For comparison, we also plot the angular
power spectrum for the purely adiabatic and isocurvature cases.  As
one can see, the CMB angular power spectrum strongly depends on
properties of the primordial density perturbations.  If there exists
correlated entropy between the baryon and other components with the
relation (\ref{Sb}), negative interference between the adiabatic and
isocurvature perturbations suppresses $C_l$ at lower multipole while
the effect of the isocurvature perturbation becomes too small to
affect the structure at high multipole.  As a result, the angular
power spectrum is enhanced at the high multipole rather than at the
low multipole.  If the effect of the entropy perturbation becomes more
efficient, then $C_l^{(\delta\phi)}$ at high multipole is suppressed
relative to that at low multipole like in the purely isocurvature
case.  This happens when the entropy perturbation is in the CDM
component with the condition given in Eq.\ (\ref{Sc}).  In addition,
with the relation given in Eq.\ (\ref{Sbc}), the correlated entropy
becomes more effective than the case where only the CDM sector has the
correlated entropy.  Then, the acoustic peaks are more suppressed
relative to the Sachs-Wolfe (SW) tail.

As mentioned before, the actual CMB angular power spectrum is given by
the sum of the inflaton contribution $C_l^{(\delta\chi)}$ and the
contribution from the primordial fluctuation of the late-decaying
scalar field $C_l^{(\delta\phi)}$.  (See Eq.\ (\ref{Cl(tot)}).)  To
parameterize their relative size, we define
\begin{eqnarray}
R_b \equiv 
S_{b\gamma}^{(\delta\phi)} / \Psi_{\rm RD2}^{(\delta\chi)},~~~
R_c \equiv 
S_{c\gamma}^{(\delta\phi)} / \Psi_{\rm RD2}^{(\delta\chi)},
\end{eqnarray}
where $S_{b\gamma}^{(\delta \phi)}$ ($S_{c\gamma}^{(\delta \phi)}$) is
the entropy between the baryon and the photon (between the CDM and the
photon) generated from the primordial fluctuation of $\phi$.  (We
adopt Eqs.\ (\ref{Sb}) and (\ref{Sc}), and hence
$S_{b\gamma}^{(\delta\phi)}$ and $S_{c\gamma}^{(\delta\phi)}$ are
equal to $\frac{9}{2}\Psi_{\rm RD2}^{(\delta\phi)}$ if they are
non-vanishing.)  The shape of the CMB angular power spectrum depends
on the values of these parameters.

Using Eqs.\ (\ref{Psi(dchi)_RD2}) and (\ref{S(dphi)}), the
$R$-parameters are given as
\begin{eqnarray}
    R_{b,c} = \frac{3}{2} 
    \left[ \frac{V'_{\rm inf}}{\phi_{\rm init}H_{\rm inf}^2} 
    \right]_{k=aH_{\rm inf}}.
\end{eqnarray}
Hence $R_{b,c}$ is model- and scenario-dependent; it depends on the
scale of inflation, shape of the inflaton potential, and initial
amplitude of $\phi$.  For example in the chaotic inflation model with
the parabolic potential $V_{\rm inf}=\frac{1}{2}m_\chi^2\chi^2$, the
above expression becomes
\begin{eqnarray}
\left[ R_{b,c} \right]_{\rm chaotic} = 
\left[ \frac{9M_*^2}{\phi_{\rm init}\chi} \right]_{k=aH_{\rm inf}}.
\end{eqnarray}
Using the fact that the inflaton amplitude at the time of the horizon
crossing of the COBE scale is $\chi\simeq 15M_*$ in the chaotic
inflation model, $\left[R_{b,c}\right]_{\rm chaotic}\simeq
0.6M_*/\phi_{\rm init}$.  Of course, the values of $R_b$ and $R_c$
depend on the model of inflation, and they vary if we consider
different class of inflation models.

\begin{figure}[t]
    \begin{center}
        \scalebox{0.75}{\includegraphics{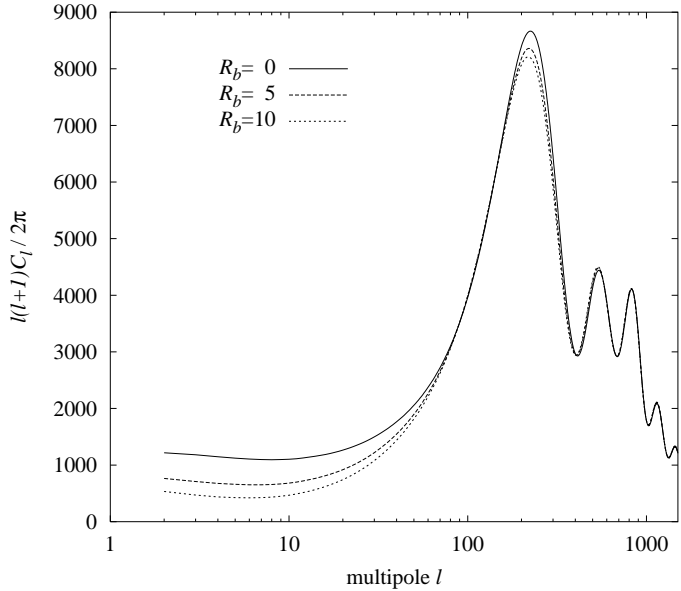}}
        \caption{The CMB angular power spectrum for the case 
        of $\alpha_b=0$.  Here we take $R_b=0$ (solid line), 5
        (long-dashed line) and 10 (short-dashed line).  The
        cosmological parameters are the same as those in Fig.\ 
        \ref{fig:Cl's}.  The overall normalizations are arbitrary.}
        \label{fig:Cl(phi)}
    \end{center}
\end{figure} 

In Fig.\ \ref{fig:Cl(phi)}, we plot the resultant angular power
spectrum with several values of $R_b$.  As expected, $C_l$ at the high
multipole is more enhanced relative to that at low ones as the
$R_b$-parameter increases.  It is important to note that the effect of
the uncorrelated entropy fluctuation always suppresses $C_l$ at high
multipole relative to that at low multipole.  Thus, if the on-going
experiments, like MAP \cite{MAP}, observes the enhancement of the
$C_l$ at high multipole, it can be regarded as a unique signal of the
late-decaying scalar condensation.  In fact, if the $R_b$-parameter is
too large, the angular power spectrum at high multipole is too much
enhanced, which becomes inconsistent with the currently available
data.  We checked that $R_b\geq 4.5$ is excluded at 95 \% C.L. even
adopting the most conservative constraint.

We also studied the effects the correlated entropy perturbation in the
CDM.  If $R_c$ is non-vanishing, $C_l$ at high multiple is suppressed
compared to the SW tails.  Thus, too large $R_c$ is also excluded from
the current data of the observations of the CMB power spectrum; $R_c$
is constrained to be smaller than $2.0$ for $R_b=0$.

In summary, we have discussed the effects of the late-time entropy
production due to the decay of the scalar-field condensations on the
cosmic density perturbations.  If the universe is reheated by the
decay of the scalar field $\phi$, many of the components in the
present universe are generated from the decay products of the $\phi$
field.  In such a case, cosmic density perturbations are affected by
the primordial fluctuation of $\phi$ which may be generated during the
inflation.

If all the components in the universe originate to the decay product
of $\phi$, density perturbations generated from the primordial
fluctuation of $\phi$ becomes adiabatic.  In this case, the CMB
angular power spectrum from the fluctuation of $\phi$ becomes the
usual adiabatic ones with (almost) scale-invariant spectrum.  If this
becomes the dominant part of the cosmic density perturbations, then we
have seen that the constraints on inflation models from observations
of the CMB angular power spectrum are drastically relaxed
\cite{MoroiTakahashi}.

If the baryon or the CDM is not generated from $\phi$, on the
contrary, correlated mixture of the adiabatic and isocurvature
perturbations may arise.  In this case, the CMB angular power spectrum
may be significantly affected and the shape of the resultant power
spectrum depends on which component has the correlated isocurvature
perturbation.  In particular, if the baryonic component has the
correlated isocurvature perturbation, $C_l$ at high multipole is more
enhanced relative to that at low multipole, which may be observed by
the MAP experiment and regarded as a unique signal of the late-time
entropy production.  

{\sl Acknowledgment:} 
This work is supported by the Grant-in-Aid for Scientific Research
from the Ministry of Education, Science, Sports, and Culture of Japan,
No.\ 12047201 and No.\ 13740138.

\end{document}